# Highly sensitive strain sensor from topological-structure modulated dielectric elastic nanocomposites

Youjun Fan, Zhonghui Shen, Xinchen Zhou, Zhenkang Dan, Le Zhou, Weibin Ren, Tongxiang Tang, Shanyong Bao, Cewen Nan, Yang Shen*

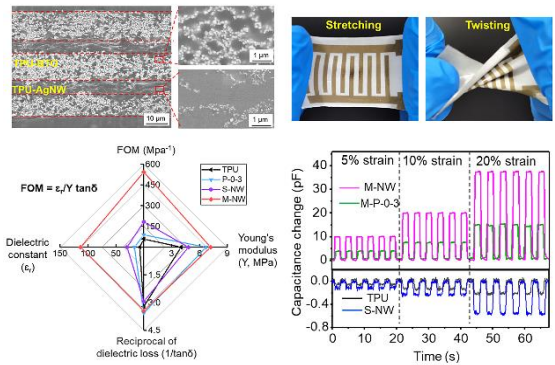

Stretchable functional materials are important components and major challenges for high-performance strain sensors. Here, a high dielectric elastic nanocomposite modulated by topological-structure is demonstrated for a flexible strain sensor. The strain sensor achieves high signal-to-noise ratio, positive sensitivity, and wide linear range. This creates a novel approach of modulating functional nanocomposite for developing high-performance flexible electronic devices.

**Keyword** dielectric elastic nanocomposites, topological structure modulation, stretchable, strain sensors

# Article

**Highly sensitive strain sensor from topological-structure modulated dielectric elastic nanocomposites**

Youjun Fan,[1,2] Zhonghui Shen,[3] Xinchen Zhou,[1] Zhenkang Dan,[1] Le Zhou,[1] Weibin Ren,[1] Tongxiang Tang,[1] Shanyong Bao,[1] Cewen Nan,[1] Yang Shen[1,2] *

[1] State Key Lab of New Ceramics and Fine Processing, School of Materials Science and Engineering, Tsinghua University, Beijing 100084, China
[2] Center for Flexible Electronics Technology, Tsinghua University, Beijing 100084, China
[3] State Key Laboratory of Advanced Technology for Materials Synthesis and Processing, Center of Smart Materials and Devices, Wuhan University of Technology, Wuhan, 430070, China
*E-mail: shyang_mse@tsinghua.edu.cn

**SUMMARY**

Flexible strain sensors are critical to several potential intelligent applications, such as human-machine interfaces, soft robotics, human motion detection, and safety monitoring of components. Stretchable functional materials are important components of strain sensors, and they are still major challenges for high performance strain sensors. Herein, we demonstrate a novel strategy of designing and optimizing flexible strain sensor by developing topological structure modulated high permittivity elastic nanocomposite. The topological structure with three-phase percolative nano-nanonetworks produces synergistic effects of space charge enhancement and local electric field modulation, and it gives rise to an ultrahigh dielectric permittivity (113.4, at 1 kHz, over 1500% enhancement than that of commercial elastic polyurethane matrix) and excellent comprehensive electromechanical performance, and the optimal comprehensive electromechanical performance reaches to 542.91 MPa$^{-1}$, which is over 9-fold than that of commercial polyurethane elastic film. An interdigital capacitive strain sensor is designed using the topological structured elastic dielectric nanocomposite. It possesses high initial capacitance density and positive capacitance response with stain, achieving high signal-to-noise ratio, high capacitance response sensitivity, and wide linear range, and it breaks through disadvantages of negative sensitivity and narrow linear range for conventional interdigital strain sensors. The prepared integrated strain sensor arrays are able to measure local strain of convoluted surfaces and monitor the motion of soft actuators in real time, and they would make conditions for intelligent control systems and the study of morphological intelligence.

**INTRODUCTION**

Flexible strain sensors sense and collect information by transducing external mechanical stimuli into electrical signals,[1,2] and they are critical to several potential applications, such as human-machine interfaces, soft robotics, human motion detection, and safety monitoring of components in aerospace.[3-5] Resistive and capacitive strain sensors are of great perspective and have been widely explored.[6,7] Resistive strain sensors depend on the resistance change of conductive materials due to the deformation of conductive network under strain. Despite of their high gauge factor, the irrecoverable deformation of conductive network usually leads to nonlinear behavior and inferior stability for strain sensors.[8-10] Capacitive sensors response mechanical stimuli by means of capacitance change, which is mainly attributed to change of geometric structure of devices (*e.g.,* effective area and thickness of the dielectric layer). Capacitive strain sensors exhibit significant advantages in linear behavior, stability, and integrated circuits.[10-12] Flexible sensors have also been fabricated by design of delicate micro-structures, such as hierarchical nano-in-micro structure, sponge-like, pyramids, micro-hairs, and fabric, and have been used in electronic skins.[13-17] While capacitive strain sensors based on structures design still exist some inherent defects, including low signal-to-noise ratio, limited sensitivity (negative values for interdigital strain sensors), and poor mechanical conformability.[14, 18-20]

Dielectric materials are essential components of flexible capacitive strain sensors and have crucial effects to their performances.[6,19] Capacitance of strain sensors are closely related to the dielectric permittivity and geometric structure of dielectric materials[19]. High permittivity of dielectric materials can provide high baseline capacitance for flexible sensors, leading to significant enhancement in signal-to-noise ratio, detection limit and sensitivity.[4] Mechanical property of dielectric materials directly affects the performance parameters and applications, such as the stretchability, linearity range, and durability.[17,18,21] Dielectric materials with high permittivity and good elasticity are highly desired for high performance strain sensors.[3,16,22] However, polymeric elastomers with good stretchability usually have ultralow dielectric permittivity. Traditional inorganic dielectric materials with high permittivity are usually rigid and brittle, restricting their applications in flexible devices.

The permittivity of dielectric materials is mainly attributed to the dipoles and interfacial space charges at low frequency electric fields.[23,24] Polymer nanocomposites incorporating with high dielectric permittivity ($\kappa$) inorganic nano-fillers have attracted much research interest for high-permittivity dielectrics.[25] A series of approaches have been reported in capacitive energy storage materials, including tailoring the interface by modifying the surface of nano-fillers, modulating anisotropy of nano-fillers, and 3D phase-field simulations.[26-28] While approaches to modulating dielectric permittivity of polymer nanocomposites have far not been applied for flexible capacitive sensors.[3] Conventional inorganic/polymer composites usually show minor enhancement in dielectric permittivity below the percolation threshold of nano-filler, which is usually very high for inorganic nano-fillers (> 30 vol% for spherical nanoparticles), resulting in high dielectric loss and significant degradation in mechanical compliance and stretchability.[23,29] The adverse coupling between dielectric permittivity and mechanical performances of conventional polymer nanocomposites poses major challenges for them as viable dielectrics in capacitive strain sensors.[24,30]

In this work, we demonstrate a novel strategy of designing and optimizing flexible strain sensor by developing topological structure modulated high permittivity elastic nanocomposite. The topological structure of nanocomposite is constructed by three-dimensional (3D) networks of nanofillers and multilayered heterogeneous films. The 3D networks formed by barium titanate (BTO) nanoparticles induces substantially enhanced electrical polarization in BTO nanocomposite layers, while silver nanowire (AgNW) networks inserted between the BTO nanocomposite layers lead to favorable concentration of local electric field in the BTO nanocomposite layers and further increase the electrical polarization. The synergistic effects of space charge enhancement and local electric field modulation induced by this topological structure give rise to an ultrahigh dielectric permittivity of 113.4 (@ 1 kHz, the same bellow, the dielectric permittivity is enhanced 1500% than that of commercial elastic polyurethane matrix), and maintains low dielectric loss of 0.029 for the elastic nanocomposite. More importantly, the nanocomposite also exhibits superior mechanical compliance and high stretchability of 360%, achieving excellent comprehensive electromechanical performance ($FOM = \frac{\varepsilon_r}{Y \cdot tan\delta}$) of 542.91 MPa$^{-1}$, which is critical to stretchable electronic devices. A flexible interdigital capacitive strain sensor is developed based on the topological-structure-modulated multilayered nanocomposite. Due to high dielectric permittivity of the nanocomposite, the flexible interdigital strain sensor exhibits high signal-to-noise ratio, high initial capacitance density, high capacitance response sensitivity, and wide linear range. Of particular interests, these strain sensors exhibit positive relation between capacitance response and the applied strain, and this performance breaks through disadvantages of the negative sensitivity and small linear range for conventional interdigital strain sensors, providing a novel approach for designing strain sensor by modulating topological structure of dielectric materials. Furthermore, the topological structured nanocomposite with high permittivity and single-side interdigital electrode make convenience for preparing integrated micro-sensor array, and a prepared integrated strain sensor array exhibits good performance in detecting the local strain of convoluted surfaces and motion states of a soft actuator.

Highly stretchable thermoplastic polyurethane (TPU) was used as polymer matrix, and commercial high-κ BaTiO$_3$ (BTO) nanoparticles and conductive silver nanowire (AgNW) were employed to construct 3D networks in TPU polymer matrix. We proposed and demonstrated that a topological-structure-modulated nanocomposite consisting of 3D networks of nanofillers and multilayered heterogeneous films could be prepared by an in-situ combinatorial fabrication process of electrospinning and electrospraying (Figure 1A).[31,32] This process makes a synchronous hybrid of polymer and inorganic nano-materials in large scale, and it creates a facile approach of micro-structure modulation for nanocomposite. TPU solution was electrospun into continuous nanofiber on a collecting aluminum film, BTO and AgNW dispersion of alcohol was electrosprayed in high voltage and assembled on the surfaces of TPU nanofiber, respectively, achieving freestanding nano-network structured nanocomposite films (TPU-BTO, TPU-AgNW films). As presented by the microstructural images in Figure 1B and 1C, BTO nanoparticles distribute on surfaces of TPU nanofiber and form aggregated chains, while AgNWs tangle on TPU nanofibers and form connected networks. Alternating layers of TPU-BTO and TPU-AgNW were assembled layer-by-layer and then hot-pressed into dense multilayered nanocomposites (noted as M-NW nanocomposite for short hereafter). Figure 1D and 1E present the three-dimensional schematic and corresponding cross-sectional microstructural images of the M-NW nanocomposite, and they show clearly networks of BTO nanoparticles and multilayer structure in the dense elastic dielectric nanocomposite. The aggregated networks of BTO and AgNW nanofillers in the alternating layers form topological structure of nanocomposite, as presented in magnified microstructural images in the right of Figure 1E. The X-ray diffraction (XRD) patterns of nanofillers (BTO, AgNW) and the energy dispersive spectrometer (EDS) images of titanium and silver elements are presented in Figure S1, confirming their presence in the final nanocomposites. The common TPU matrix among the alternating layers gives rise to dense cross section and less defects on interfaces after hot-pressing, which ensures excellent mechanical compliance and stretchability, as shown by the picture of Figure 1F. The M-NW nanocomposite possesses high stretchability of ~ 360%, and it is superior than single layered S-NW nanocomposite and TPU-AgNW composite (Figure 1G). The strain cycling behavior of the M-NW nanocomposite film was also measured and presented in Figure 1H. The dynamic mechanical properties of these nanocomposites were characterized by DMA (Figure S2A, B). The storage modulus of the M-NW nanocomposite is larger than that of the S-NW nanocomposite and TPU-AgNW nanocomposite at low temperatures.

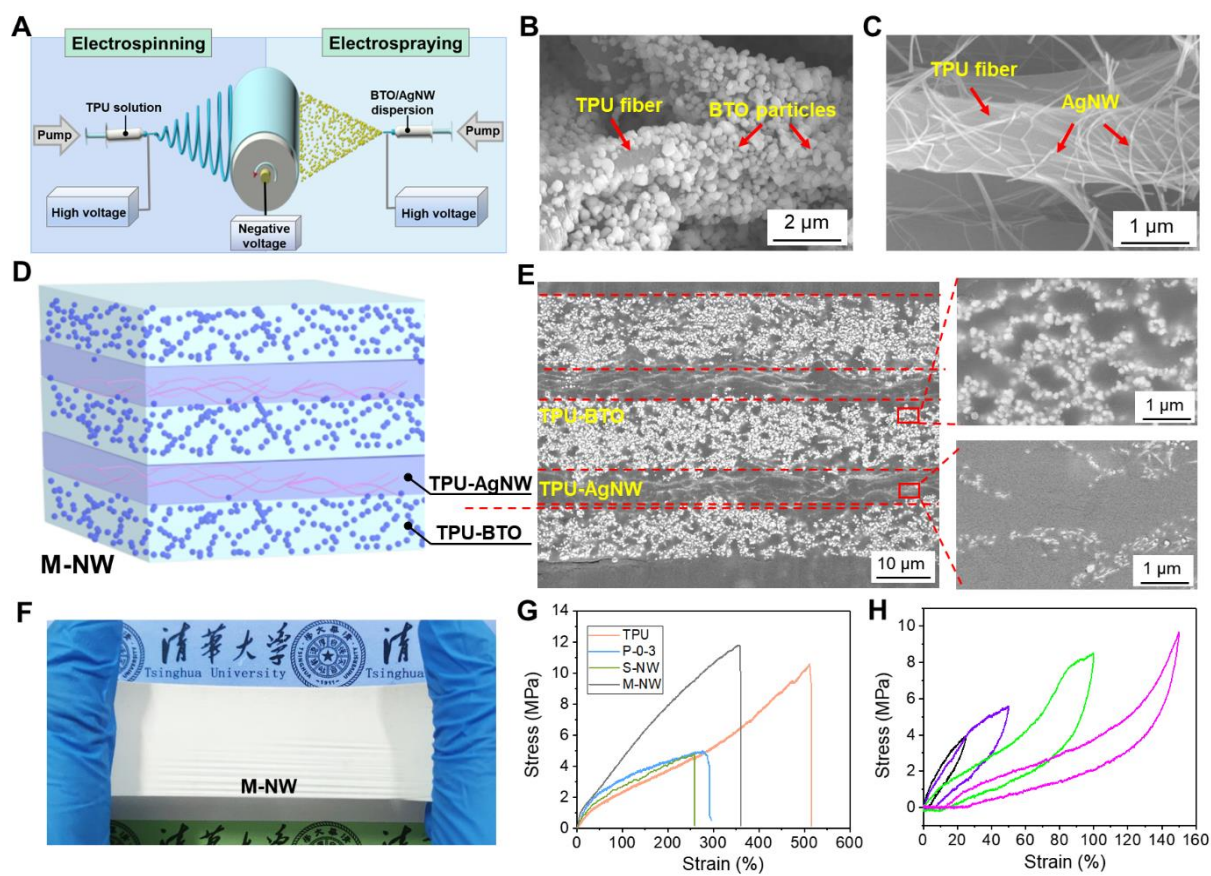

**Figure 1. Fabrication process and topological structure of the elastic nanocomposite.** (A) Schematics of the in-situ combinatorial fabrication process. (B, C) SEM images of TPU-BTO (B) and TPU-AgNW (C) nanofibers of fibrous nanocomposite films prepared by the in-situ combinatorial fabrication method. (D) The micro-structure schematic of multilayered topological structured nanocomposite (M-NW nanocomposite). (E) Cross-sectional view of the M-NW nanocomposite, presenting overlapped TPU-BTO layers and TPU-AgNW layers, and magnified profile images of BTO and AgNW nanofillers in the right images. (F) Optical photograph of the stretched M-NW nanocomposite film. (G) Stress–strain curves of TPU film and different structured nanocomposites (P-0-3, S-NW, M-NW). (H) Strain cycling behavior of the freestanding M-NW nanocomposite film.

For comparison, nanocomposites with BTO nanoparticles homogeneously dispersed in TPU matrix (P-0-3) and with only aggregated BTO networks but no AgNWs (S-NW) were also prepared with identical process. The dielectric permittivity of S-NW nanocomposite increases with the increasing of the diameter and content of BTO nanoparticles (Figure 2A, 2B), due to the fact that lager BTO nanoparticles are more easily aggregated to form connected network. Figure 2C present the three dimensional structures and corresponding cross-sectional micro images of different nanocomposites, and they are dense and voids-free. The most striking feature to Figure 2E is the distinctive effects of topological structure on the dielectric behavior of the TPU-based nanocomposites. By simply confining BTO nanoparticles into 3D nanonetworks (as in S-NW nanocomposite) instead of dispersing them randomly in the TPU-matrix (as in P-0-3 nanocomposite), we achieve substantially enhanced dielectric permittivity for the S-NW nanocomposite ($\varepsilon_r \sim 30$) compared to the P-0-3 counterpart with $\varepsilon_r$ of ~ 17 (at 1 kHz) at the same content of BTO nanoparticles. Further assembly of the alternating S-NW and TPU-AgNW layers into topological-structure-modulated M-NW nanocomposite gives rise to even much higher dielectric permittivity. An ultrahigh dielectric permittivity of 113.4 is obtained for the M-NW nanocomposite at a low loading of 11 vol.% BTO nanoparticles, which is an enhancement of ~ 1500% over than that of pure TPU ($\varepsilon_r \sim 7.3$). More importantly, the M-NW nanocomposite also exhibits rather low dielectric loss of ~ 0.029 within a broad frequency range, indicating that the nanocomposites remain highly insulating. To our best knowledge, it is by-far the best comprehensive dielectric performances ever achieved for elastic dielectric nanocomposites (Table S1). Although even higher dielectric permittivity has been reported in percolative composites filled with carbon nanotubes (CNTs), the rather high dielectric loss induced by the percolating paths formed by CNTs may result in substantially increased leakage current and prevent them as viable dielectrics for strain sensors.[33,34] By contrast, the multilayered nanocomposite constructed by pure TPU and TPU-AgNW layers shows low dielectric permittivity (Figure S3), hence the dielectric properties of the multilayered nanocomposites do not simply rely on the average effects of their constituent layers. Results of phase-field simulations suggest that the largely enhanced dielectric permittivity of the M-NW nanocomposite could be mainly attributed to the networks of aggregated BTO nanoparticles and heterogeneous interfaces between the alternating BTO and AgNW layers. Firstly, aggregation of BTO nanoparticles into 3D networks leads to much pronounced concentration of local electric field between the neighboring BTO nanoparticles, as indicated by the intensified red color in the distribution of local electric field simulated by phase-field models shown in Figure 2D. Therefore, BTO nanoparticles are subjected to higher local electric field and exhibit enhanced electrical polarization at a low external electric field.[30,35] Secondly, the TPU-AgNW layers with higher electrical conductivity (the AC conductivity as a function of frequency of nanocomposites are shown in Figure S4) may further distort the distribution of electric field among the neighboring layers. Calculation of local electric fields distribution by the principle of capacitive voltage divider indicates that even more serious concentration of local electric field in the BTO nanoparticle layers (presented in Figure S5) is induced by the multilayer structure.[24] Higher local electric field in the BTO layers is then favorable for inducing higher electrical polarization of the BTO nanoparticles. Thirdly, due to the large differences in dielectric permittivity and electrical conductivity of TPU-BTO layers and TPU-AgNW layers, the interfaces between the neighboring layers may serve as traps and additional charge accumulation sites for the M-NW nanocomposite, leading to much enhanced interfacial polarization hence higher dielectric permittivity.[23] In addition to the enhanced interfacial polarizations, the heterogeneous interfaces also give rise to suppressed leakage current of the M-NW nanocomposite at high electric field, as evidenced by the electric polarization-electric field (P-E) loops (shown in Figure S6). For instance, after polarized at an electric field of 60 MV/m, the remnant polarization of the M-NW nanocomposite is only ~ 0.9 μC/cm$^2$ compared to a remnant polarization of ~ 2.7 μC/cm$^2$ for the S-NW nanocomposites with the absence of heterogeneous interfaces among layers. The pronounced interfacial polarization in M-NW nanocomposite could also be distinguished in the variation of dielectric permittivity with temperature (Figure S7), as indicated by the dramatical increase of dielectric permittivity with temperature in the low frequency range.

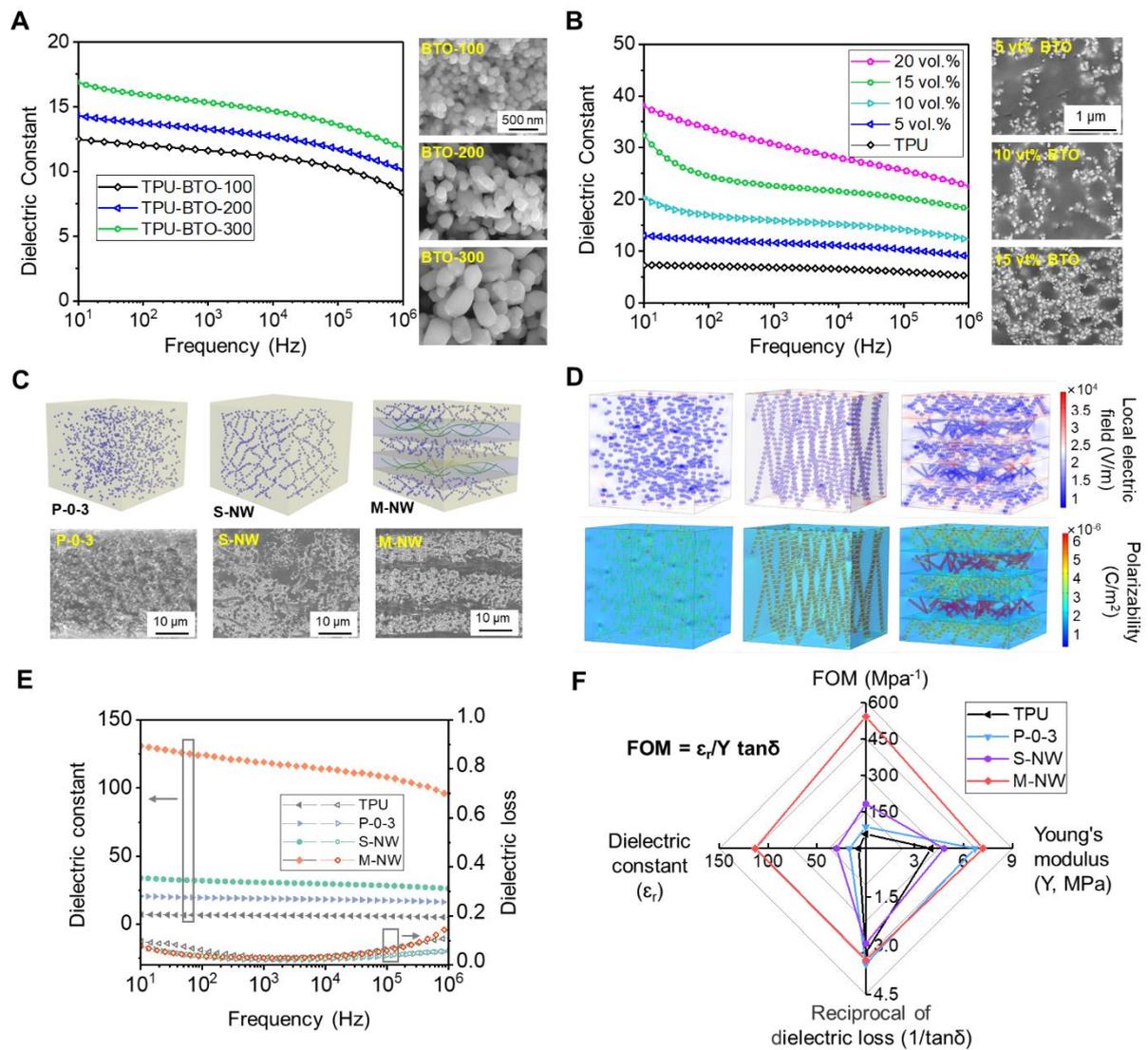

**Figure 2. Dielectric property and comprehensive electromechanical performances of different structured nanocomposites.** (A,B) Frequency dependences of dielectric permittivity of S-NW nanocomposites with different BTO diameter (A) and different BTO content (B), and the right figures show their SEM images. (C) Micro-structure schematics and corresponding micro-structure images (cross-section) of three kinds nanocomposites (P-0-3, S-NW, and M-NW). (D) The frequency dependences of dielectric permittivity and dielectric loss for TPU and corresponding nanocomposites. (E) Simulated electric field and polarizability distribution by phase-field method for corresponding different structured nanocomposites. (F) Comprehensive electromechanical performance of TPU and corresponding nanocomposites.

For flexible high-performance strain sensors, comprehensive electromechanical performances, *e.g.*, high dielectric permittivity, low dielectric loss and low Young's modulus, are mandatory for the dielectric media. We therefore propose and define a comprehensive electromechanical figure of merit as FOM = $\frac{\varepsilon_r}{Y \cdot \tan\delta}$, where $\varepsilon_r$ and $\tan\delta$ are relative dielectric permittivity and dielectric loss of nanocomposites, and Y is the Young's modulus of nanocomposite. The dielectric and mechanical properties of a series of nanocomposites are summarized in Table S2. We then compare FOMs of TPU and the corresponding different nanocomposites in the radar chart in Figure 2F. The comprehensive electromechanical performance of M-NW nanocomposite reaches to 542.91 MPa$^{-1}$, which is over 9-fold than that of pure TPU elastic film.

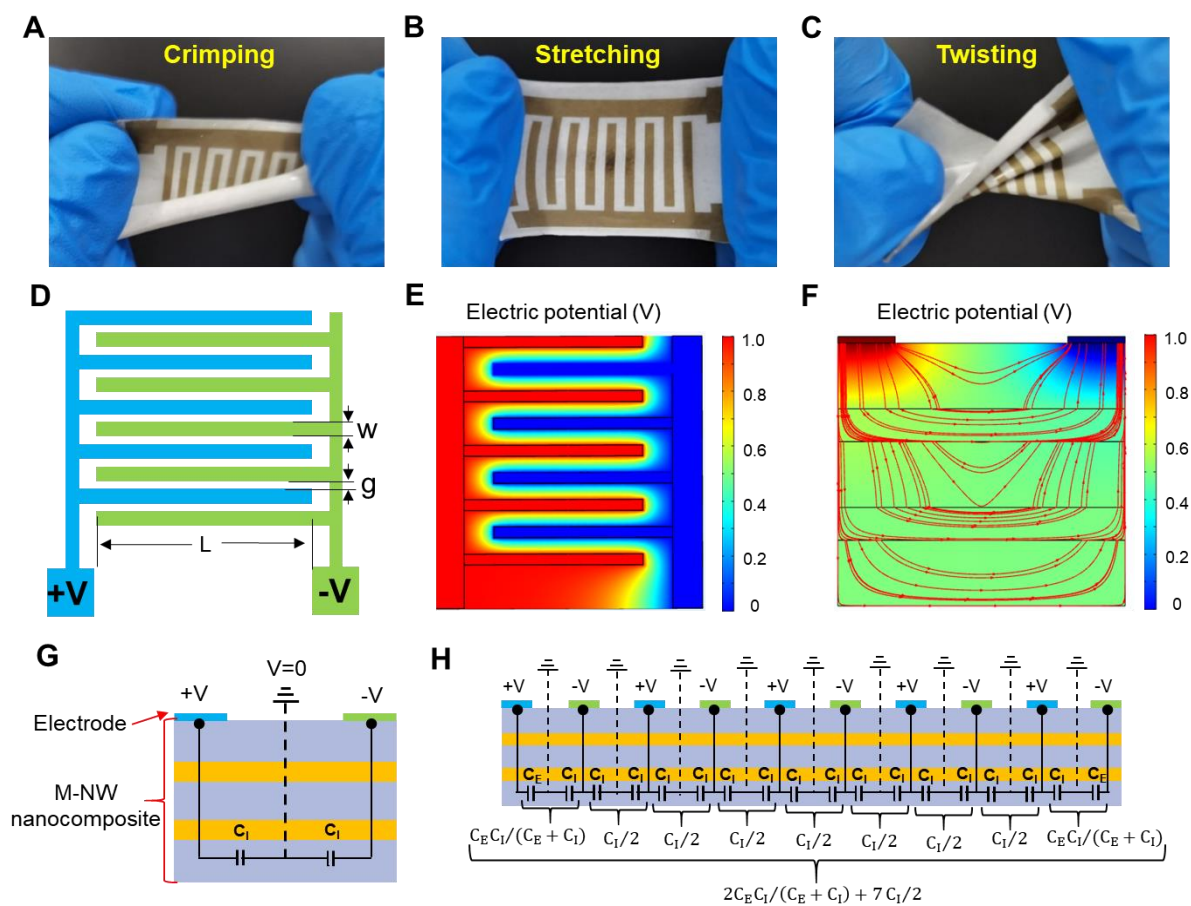

**Figure 3. M-NW nanocomposite-based interdigital strain sensors and the local electric potential distribution.** (A-C) Photographs of the crimped, stretched and twisted interdigital strain sensors, showing excellent flexibility. (D) The structure of the top side interdigital electrode. (E, F) Electric potential of top view (E) and magnified local side view of the M-NW nanocomposite between neighbor electrode fingers (F) for the interdigital strain sensor with 1 V applied. (G, H) Scheme for the capacitance of a capacitor unit between neighbor electrode fingers (G) and the whole interdigital strain sensor (H).

A highly conductive AgNW-based nanocomposite film was applied as the electrode for fabricating the strain sensor,[31] and it possesses excellent mechanical compliance and stretchability (as shown in Figure S8). The stretchable conductive film was designed and prepared into interdigitated elastic electrode by laser scribing process (the photograph and micro-structure of the electrode are presented in Figure S9A, B). A flexible interdigital capacitive strain sensor was fabricated by constructing the interdigitated electrode with the M-NW nanocomposite. The interdigital strain sensor is highly flexible with low Young's modulus of ~ 2.2 MPa (Figure S10). It can also sustain various mechanical deformations (Figure 3A-3C). The interdigital strain sensor makes integration of several parallel partial capacitance due to the interdigital electrode (Figure 3D). Figure 3E and 3F present the COMSOL calculated top view electric potential and magnified local side view electric potential of M-NW nanocomposite between neighbor electrode fingers.[36] The perpendicular plane halfway between neighbor electrode fingers is equipotential plane with V = 0, and electric field lines cross these equipotential planes, as illustrated by the equivalent circuit of per capacitor unit between neighbor electrode fingers in Figure 3G. The analytical model and calculating formulas of the integrated capacitance is presented in Figure 3H, and the total capacitance of the interdigital strain sensor is calculated by the follow equation: [37,38]

$$C = (N-3)\frac{C_I}{2} + \frac{2C_E C_I}{C_E + C_I}, \text{ for } N > 3 \tag{1}$$

where N is the number of electrode fingers; $C_I$ is half the capacitance of one interior electrode relative to the ground potential, and $C_E$ is the capacitance of one outer electrode relative to the ground plane. The detailed calculation is illustrated in Supporting Information. The M-NW nanocomposite-based interdigital strain sensor exhibits the largest initial capacitance density of 31.41 nF/cm$^3$, owing to the high dielectric permittivity of dielectric layer, and it is 125.6 and 61.6 times than that of pure TPU and S-NW nanocomposite-based interdigital strain sensors (Figure 4A). This can translate into much increased signal-to-noise ratio for the strain sensor especially under small strain. The capacitance usually decreases under longitudinal strain for conventional interdigital strain sessors, which is mainly attributed to the increase of gap distance between the electrode fingers. As indicated by the negative capacitance response curves of TPU and S-NW nanocomposite based interdigital strain sensors (black and blue curves in Figure 4B and Figure 4C). This feature leads to negative gauge factor and small linear range, limiting the applications of interdigital strain sensors. On the contrary, the M-NW nanocomposite-based interdigital strain sensor presents positive capacitance response with increasing strain, as shown by the pink lines in Figure 4B and 4C and Supporting Video S1, and this performance gives rise to much higher capacitance response sensitivity ($\frac{\Delta C}{\varepsilon}$) of 5.7 pF/ % (corresponding to a gauge factor ($\frac{\Delta C/C}{\varepsilon}$) of 1.06) and wide linear range (from 0.1% to 100%, Figure S11). It exhibits superior performance than most reported interdigital strain sensors, and comparisons among different capacitive strain sensors are presented in the Supporting Information (Table S3). The change of capacitance under strain for the interdigitated capacitor is mainly related to the change in the properties and electric field of the deformed dielectric materials. This is presented in the equation:[39]

$$\frac{\Delta C}{C} = \frac{\Delta \varepsilon}{\varepsilon} + \frac{\Delta E}{E} \qquad (2)$$

where $\varepsilon$ is the dielectric permittivity of the dielectric material, and $E$ is the electric field near electrodes. The normal deformation of dielectric layer affects the change of dielectric constant and electric field, as shown in equations:

$$\frac{\Delta \varepsilon}{\varepsilon} = K_n^\varepsilon \frac{\Delta d}{d} \qquad (3)$$

$$\frac{\Delta E}{E} = K_n^E \frac{\Delta d}{d} \qquad (4)$$

where $d$ is the thickness of dielectric layer, and parameter $K_n^\varepsilon$ and $K_n^E$ are defined by the materials composition and can be theoretically estimated.[40,41] The total capacitance response for a normal deformation is

$$\frac{\Delta C}{C} = (K_n^\varepsilon + K_n^E) \frac{\Delta d}{d} \qquad (5)$$

Therefore, the capacitance response with strain is relate to the structure and normal deformation of the dielectric layer. Due to the multilayered structure and the concentration of local electrical field,[42] the M-NW nanocomposite possess positive coefficient parameters of dielectric constant and electric field near electrodes under strain.

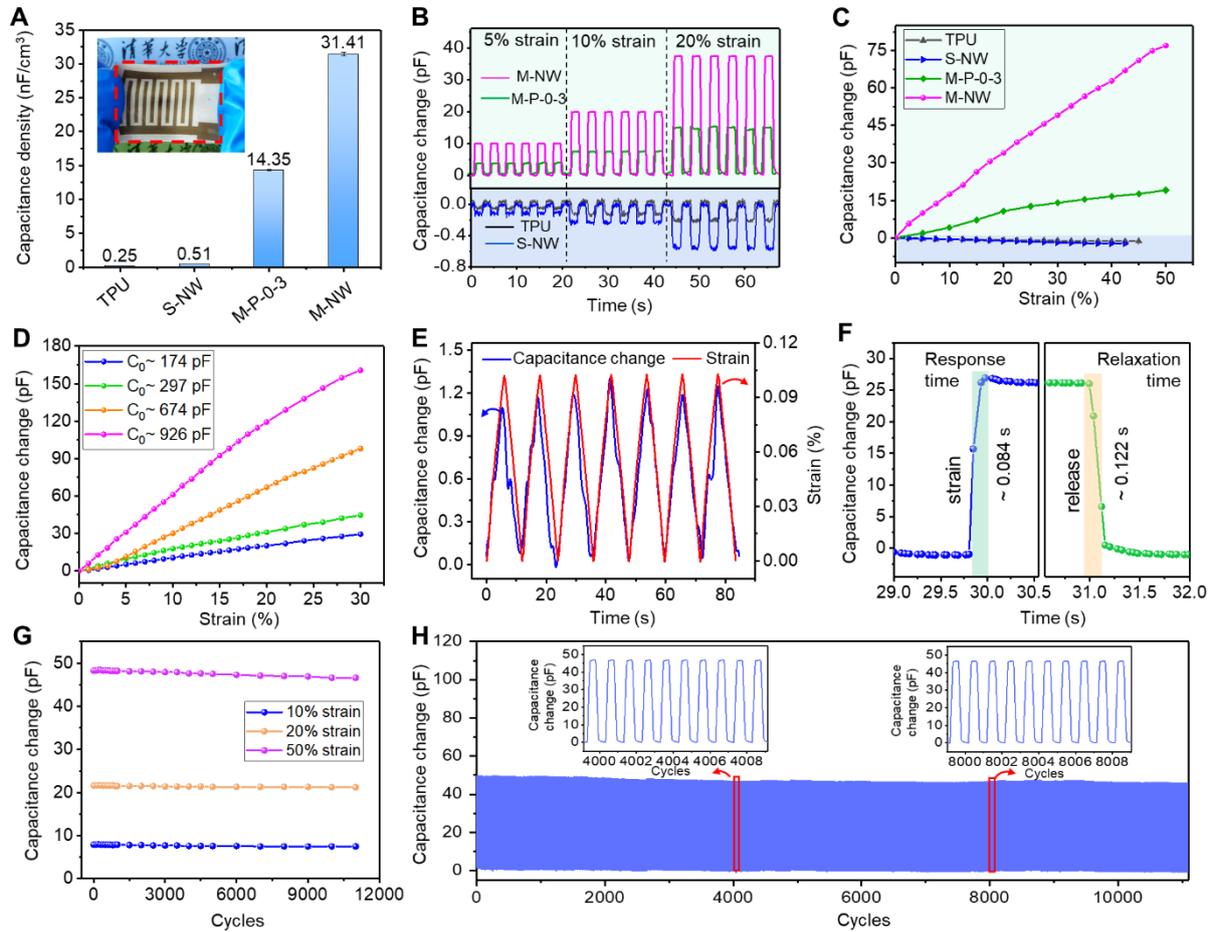

**Figure 4. Performance of flexible interdigital strain sensors based on topological-structured nanocomposites.** (A) Capacitance density of interdigital strain sensors from different structured nanocomposites, and the inset show the photograph of an interdigital strain sensor. (B, C) Time-resolved capacitance change response of these interdigital strain sensors under cyclic external strain (B), and corresponding sensitivity curves of capacitance change (C). (D) Sensitivity of capacitance change for the M-NW nanocomposite-based strain sensor with different initial capacitance. (E) Relative capacitance change (blue line) under cycling strain (the maximum 0.1% strain, red line) for the M-NW nanocomposite-based interdigital strain sensor. (F) Dynamic response and recovery time for the interdigital strain sensor. (G) Stability of the interdigital strain sensor as a function of over 10,000 cycles under different cyclic strain. (H) Stability of the strain sensor as a function of over 11,000 cycles of 50% strain.

For the M-NW nanocomposite-based interdigital strain sensors, the linear capacitance response *vs* strain curves are related to their initial capacitance (Figure 4D). Larger initial capacitance corresponds to higher capacitance change sensitivity, and the maximum sensitivity ($\frac{\Delta C}{\varepsilon}$) reaches to 5.53 pF %$^{-1}$ (the initial capacitance of ~ 926 pF). Large initial capacitance of the strain sensor also produces ultralow detection limit, as presented in Figure 4E, the capacitance change response curve (blue line) is closely related to the strain curve (red line) under the cyclic strain of ~ 0.1%. Also, attributing to the good elasticity of M-NW nanocomposite dielectric layer and excellent compatibility with electrode, the strain sensor exhibits fast response time of 0.084s and 0.122s corresponding to the loading and unloading strain, respectively (Figure 4F). The strain sensor possesses excellent robustness under cyclic deformations, as presented in Figure 4G, it maintains initial performance even after 11000 cycles continuous cyclic stretching of 10%, 20%, 50%, (the cyclic curves of capacitance change are presented in Figure 4H, a strain sensor with initial capacitance of ~ 297 pF is measured), this indicates the capability of long-term applications for the strain sensor.

Because of the integrated and stretchable structure, the strain sensor is capable to sense various deformations and produces different capacitance response. A series of deformations of longitudinal strain, transverse strain and bending were measured for the strain sensor, and the normalized capacitance change response curves are presented in Figure 5, respectively. When applied longitudinal strain, the strain sensor produces dramatically capacitance change, and it has the highest sensitivity (Figure 5A-5C). While, the capacitance response sensitivity in transverse strain (Figure 5D-5F) and bending deformation (Figure 5G-5I) is lower than that of longitudinal strain. This different capacitance change sensitivity is capable to distinguish types of deformations by mathematical calculation. In application, the strain sensor can be directly used in the detection of various body motions in real time (Figure S12).

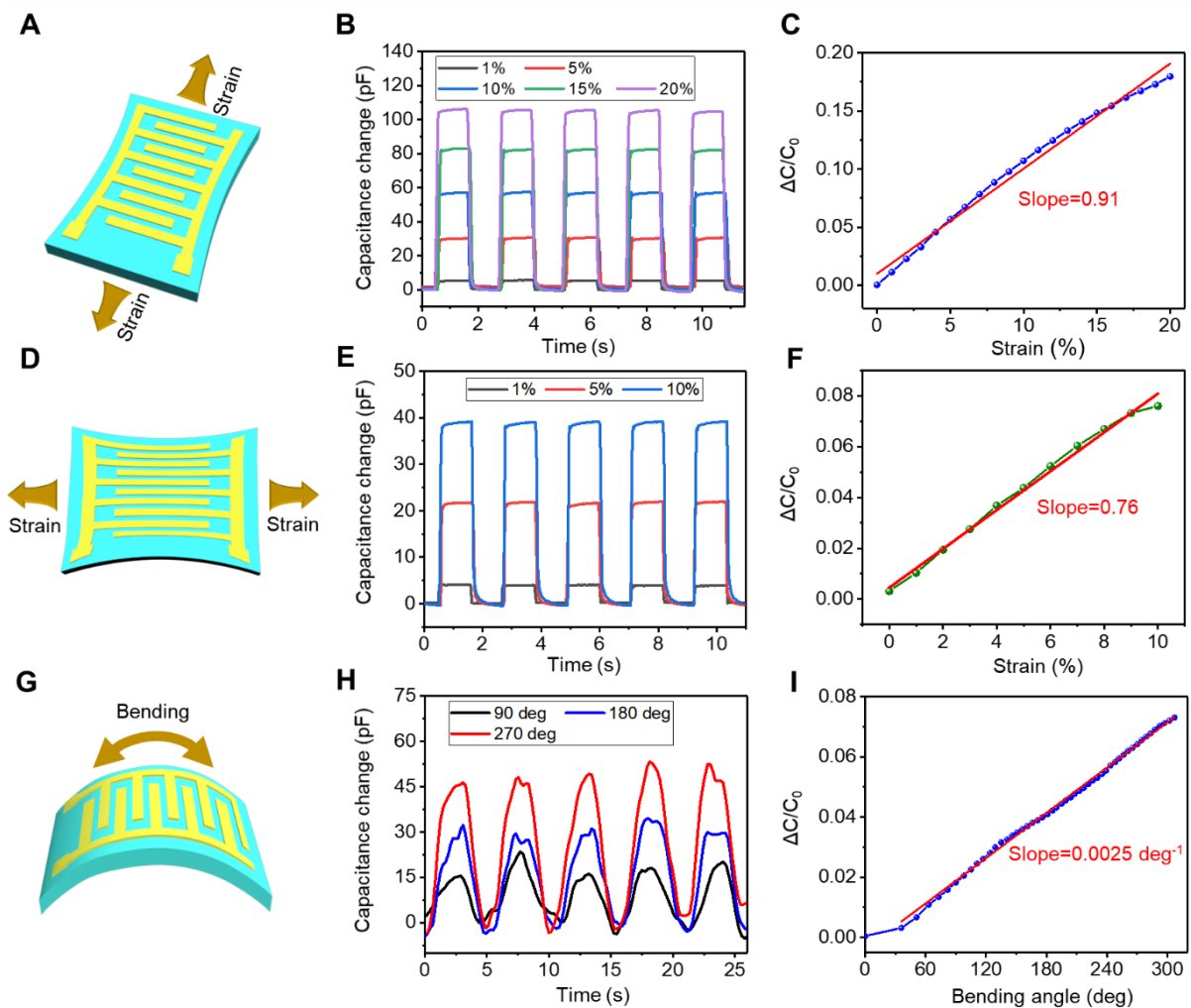

**Figure 5. Detection for various mechanical strain by the M-NW nanocomposite-based interdigital strain sensor.** (A-C) Cyclic capacitance response and normalized linear sensitivity as a function of longitudinal strain. (D-F) Cyclic capacitance response and normalized linear sensitivity as a function of transverse strain. (G-I) Cyclic capacitance response and normalized linear sensitivity as a function of bending.

Detecting the local deformation of components is an important feature of strain sensors, and integrated strain sensor arrays are highly desired. High permittivity of the M-NW nanocomposite and single-sided electrode configuration provides conditions for fabricating integrated strain-sensor arrays. A stretchable integrated strain sensor array model with micro-pixels is demonstrated for position sensing (Figure 6A). We created a 4 × 4 pixels integrated strain-sensor array (ISSA) with a pixel area of 5 mm × 5 mm (presented on sphere of a balloon in Figure 6B), and the maximum initial capacitance of a strain-sensor pixel reaches about 48.9 pF. The ISSA can conformally bond to various curvilinear surfaces and expandable deployments, and detect the deformation of local regions by capacitance response of every pixel. As presented in Figure 6C, the strain sensor array measured the strain distribution of the film as pocking by a bar, presented by the mapping capacitance response in Figure 6D, and the maximum strain is about 14.4% calculated by the capacitance response. Furthermore, an integrated system of actuating and sensing is developed by integrating a pneumatic soft actuator and a strain sensor array (Figure 6E). The flexible actuator was constructed by a series of chambers in parallel, with an integrated strain sensor array inserting on its backside, and the motion state of the soft actuator can be detected by the strain sensor array in real time. The flexible actuator moves due to the expansion of chambers as blowing gas,[43] and the deformation degree of the soft actuator in different regions is presented by capacitance change of every pixel.

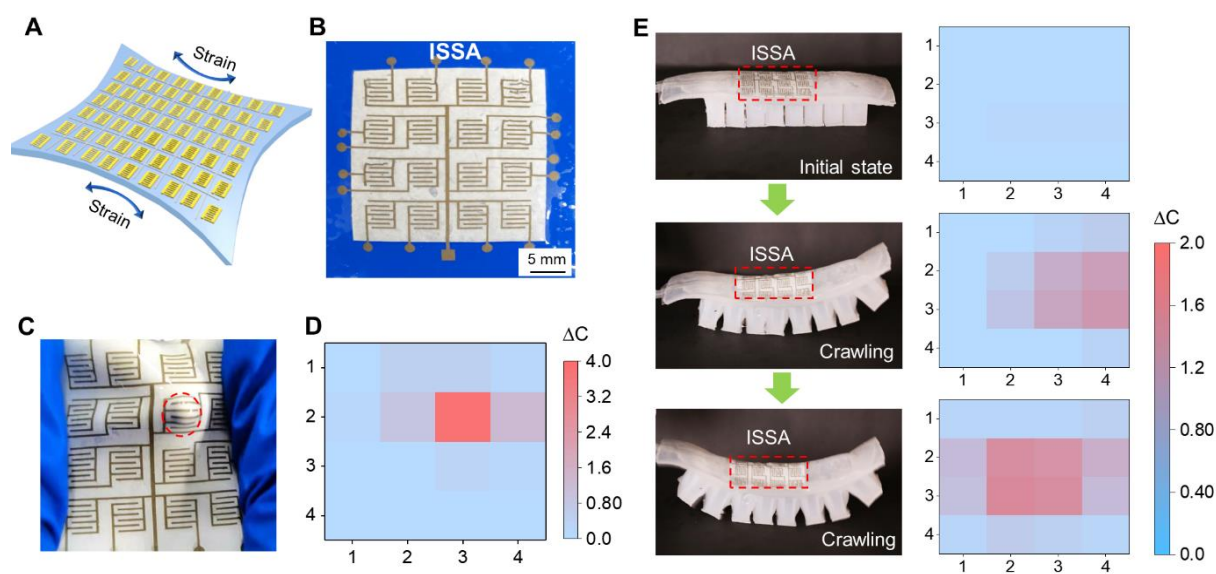

**Figure 6. The integrated strain sensor array (ISSA) and its performance in regional deformation detection.** (A) The schematic of an integrated strain sensor array. (B) Photograph of a 4 × 4 integrated strain-sensor array on an expanded balloon. (C, D) Photographs of the integrated strain-sensor array as poked by a bar and corresponding capacitance change response mapping. (E) An integrated sensing and actuating device and its three different motion states as moving, and the right corresponding capacitance change response mappings present deformation degree of every local region by the ISSA.

Three motion states for the moving soft actuator are depicted in Figure 6E, and corresponding digital plots of capacitance change for every pixel of strain sensor array present these motion states (the right in Figure 6E). This technique may be potential for the sensing and controlling system of soft robots and intelligent expandable deployment in aerospace.

In summary, an elastic nanocomposite (M-NW nanocomposite) with high permittivity and good stretchability is developed based on the topological structure modulation. This topological structure consists of 3D nanofillers networks and multilayered heterogeneous films, which produces synergistic effects of space charge enhancement and local electric field modulation, achieving a high dielectric permittivity of 113.4 and excellent mechanical properties for the nanocomposite. Based on this elastic nanocomposite, a novel flexible interdigital capacitive strain sensor is demonstrated, presenting positive response relation between capacitance change and the applied strain, and it is differing from conventional typical interdigital capacitive strain sensors. The interdigital strain sensor has high signal-to-noise ratio, high sensitivity and wide linear range. The M-NW nanocomposite makes conditions for preparing integrated micro-point sensor arrays, and an integrated system of actuating and sensing is demonstrated, achieving autonomous detection for its motion states. This work reveals a potential approach of developing high performance sensing system by modulating the topological structure of functional nanocomposites.

**EXPERIMENTAL PROCEDURES**

**Fabrication of fibrous nanocomposite films**

TPU particles (Elastollan, 1180 A) were dissolved in the solvent of hexafluoroisopropanol (HFIP, Aladdin) at concentration of 4 wt% (weight ratio). After magnetic stirring for about 6 hours at room temperature, a stable TPU solution was obtained. BTO nanoparticles were dispersed in alcohol at concentration of 0.2 g/ml with aid of ultrasonic treatment, and AgNW was dispersed in alcohol at concentration of 1 mg/ml. An electrospinning device with high voltage systems (Ucalery, ET-2535H) was used for the in-situ combinatorial fabrication process. TPU nanofibers were electrospun under a positive voltage of 10 kV, and BTO nanoparticles were electrosprayed at the other end connected with 9 kV high voltage, simultaneously, and they were collected on a shared aluminum foil wrapped around a revolving roller, connecting with a negative voltage of 2 kV. BTO nanoparticles assembled on the surfaces of TPU nanofibers, and the fast evaporation of solvent makes effective connection between nanoparticles and TPU nanofibers. A fibrous composite film of TPU-BTO was prepared and it was peeled from the aluminum foil with the help of a PET (Polyethylene terephthalate) frame. The similar process was employed to fabricate fibrous TPU-AgNW film.

**Fabrication of M-NW nanocomposite**

Three pieces of TPU-BTO fibrous films and two pieces of TPU-AgNW films were stacked layer by layer, and they were hot-pressed at 150 °C for 120 min under pressure of 5 MPa. A dense M-NW nanocomposite film was obtained and cold in the air. The individual S-NW nanocomposite and dense TPU-AgNW films were prepared using the similar hot-pressing process.

**Fabrication of interdigital capacitive strain sensor**

A TPU-AgNW elastic electrode was firstly prepared by the in-situ combinatorial fabrication process. The elastic electrode was laminated on the as-prepared dielectric nanocomposite, and the designed pattern of the interdigitated electrode was cut by laser in a suitable power. The flexible interdigital strain sensor was fabricated by the above method and then two conductive wires connected at two ends of electrode by silver paste, and the flexible strain sensor was then packaging by PDMS (polydimethylsiloxane).

**Measuring of the strain sensor**

The strain sensor was installed on a stepping motor and was applied accurate uniaxial strain. The capacitance of the strain sensor at different strain was measured with a HP 4980A precision impedance analyzer (Agilent Technologies Inc.), and the measured capacitance value was recorded by impedance testing software synchronously. The mathematic relationship between the capacitance and applied strain was then calculated.

**Fabrication of an integrated sensing and actuating device**

A pneumatic soft actuator was prepared by a 3D printing mold firstly. Ecoflex 00-30 A and B components were mixed well at ratio 1:1 by volume, and then removed bubbles by vacuumizing. The Ecoflex was added into the mold and cured at 80 °C for 1 hours, then peel off the Ecoflex chambers mold and connected an air tube to form the top half of the pneumatic soft actuator. A flat PDMS film adhered with an integrated strain sensor array was prepared, and it was adhered on the bottom of the Ecoflex chambers mold by uncured PDMS, thus an integrated sensing and actuating device was prepared.

**ACKNOWLEDGMENTS**

This research was supported by the National Science Foundation of China (Grant No. 51625202 & 52002203), and Chinese Postdoctoral Science Foundation (Grant No. 2020 M 670316). We thank Prof. Guang Zhu, Dr. Jian Chen and Pengtao Yu (Beijing Institute of Nanoenergy and Nanosystems, Chinese Academy of Sciences) for their help with the fabrication of flexible devices.

**DECLARATION OF INTERESTS**

The authors declare no conflict of interest.

# Supporting Information

**Highly sensitive strain sensor from topological-structure modulated dielectric elastic nanocomposites**

*Youjun Fan, Zhonghui Shen, Xinchen Zhou, Zhenkang Dan, Le Zhou, Weibin Ren, Tongxiang Tang, Shanyong Bao, Cewen Nan, Yang Shen\**

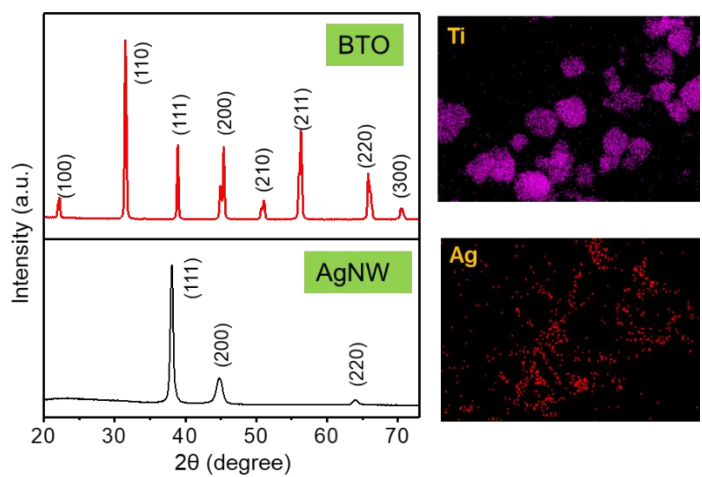

**Figure S1.** XRD patterns for BTO nanoparticles and silver nanowire (AgNW), and EDS images of titanium and silver elements.

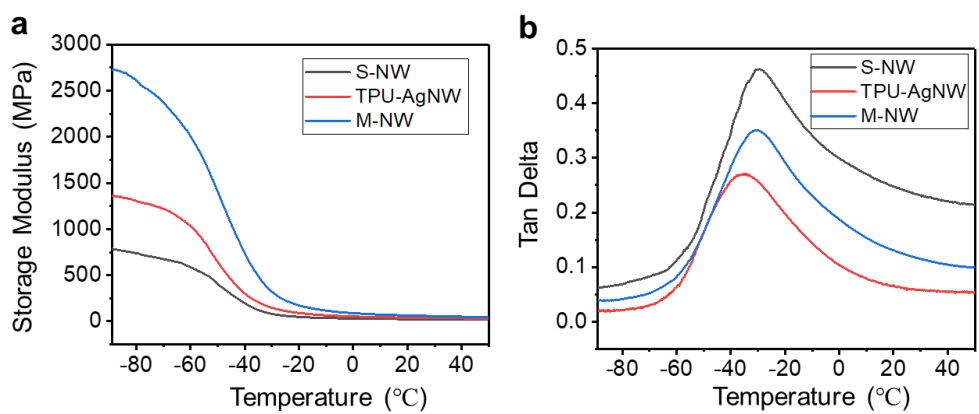

**Figure S2.** Dynamic mechanical analysis (DMA) of S-NW nanocomposite, TPU-AgNW nanocomposite and M-NW nanocomposite films.

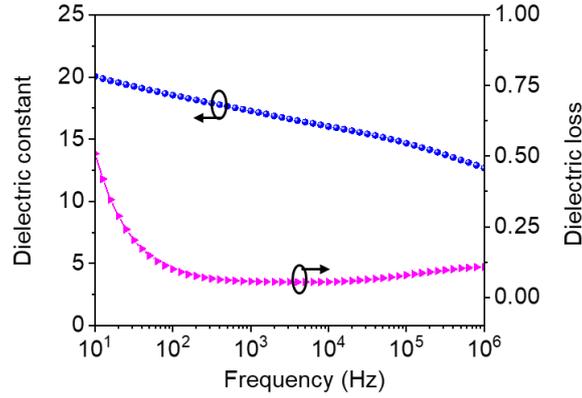

**Figure S3.** Frequency dependences of dielectric permittivity and dielectric loss of multilayered nanocomposite constructed by pure TPU and TPU-AgNW layers.

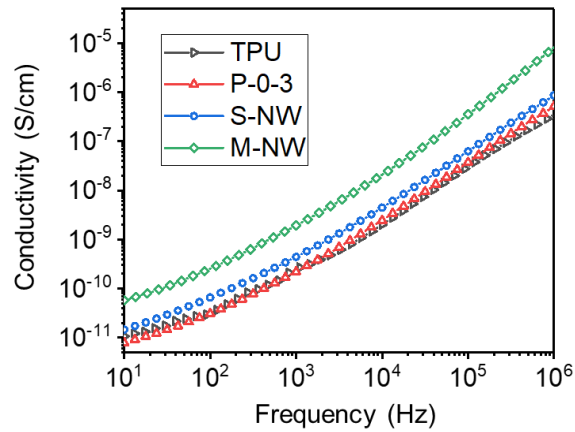

**Figure S4.** Frequency dependences of conductivity of TPU and different structured nanocomposites (P-0-3, S-NW, M-NW).

## Calculations for local electric fields of the M-NW nanocomposite

Local electric fields of the S-NW nanocomposite are calculated according to the principle of capacitive voltage divider. The five constituent layers can be considered as five capacitors in serial connection, thus the equivalent capacitance equal to

$$C_{eq} = \frac{1}{2 \times \frac{1}{C_A} + 3 \times \frac{1}{C_B}} \quad (1)$$

where $C_A$ and $C_B$ donate the divide capacitance of TPU-AgNW layer (A-layer) and TPU-BTO layer (B-layer), respectively. According to the principle of capacitive voltage divider, the local electric field for individual layer is calculated by following equations:

$$E_A = \frac{V}{2d_A + 3\frac{\varepsilon_A}{\varepsilon_B}d_B} \tag{2}$$

$$E_B = \frac{V}{2\frac{\varepsilon_B}{\varepsilon_A}d_A + 3d_B} \tag{3}$$

Where $d_A$ and $d_B$ are the thickness of A-layer and B-layer, and $\varepsilon_A$ and $\varepsilon_B$ are the relative dielectric permittivity of A-layer and B-layer, respectively. Due to $\varepsilon_A \gg \varepsilon_B$, the local electric field of B-layer ($E_B$) is dramatically increased.

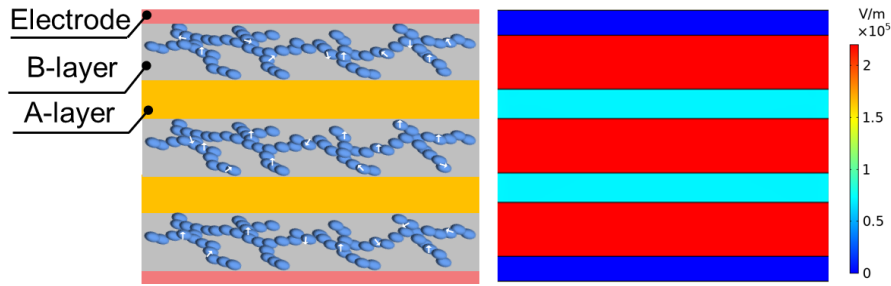

**Figure S5.** Schematics of cross-sectional structure for the M-NW nanocomposite and corresponding calculated local electric fields distribution of component layers as applied voltage.

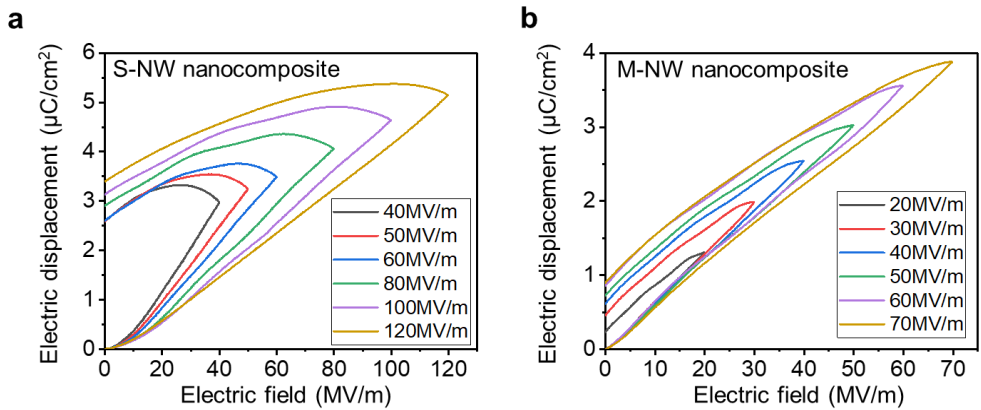

**Figure S6.** Electric polarization-electric field (P-E) loops for S-NW nanocomposite (**a**) and M-NW nanocomposite (**b**).

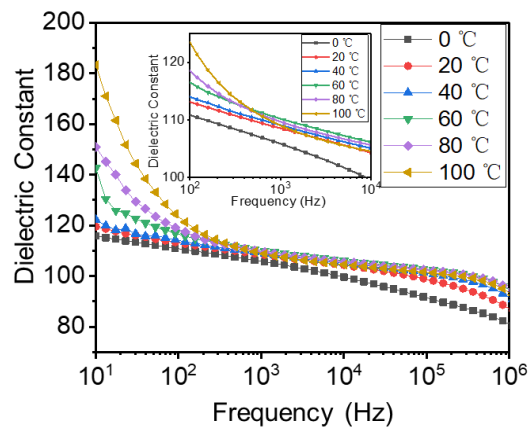

**Figure S7.** Dielectric permittivity as a function of frequency at various temperatures for M-NW nanocomposite.

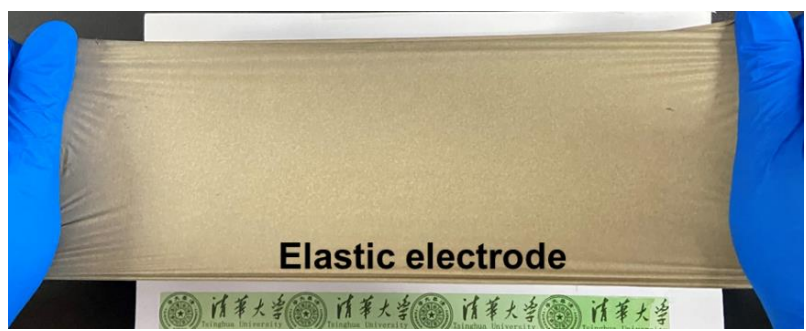

**Figure S8.** Photograph of AgNW-based elastic electrode.

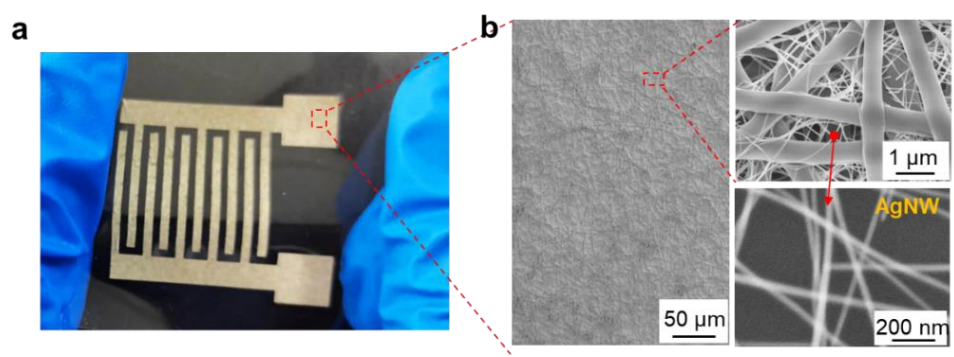

**Figure S9.** Photograph (**a**) and micromorphology images (**b**) of interdigital AgNW-based elastic electrode.

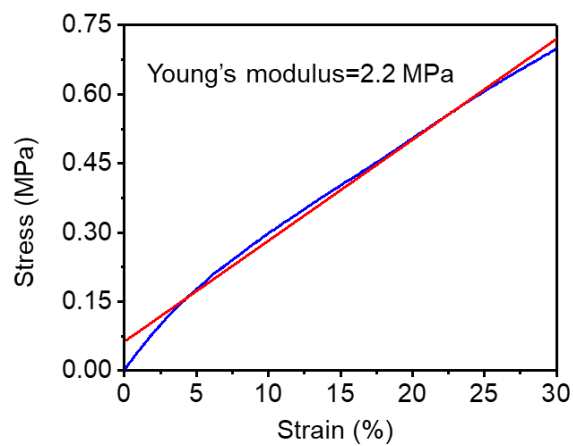

**Figure S10.** Stress–strain curve of the M-NW nanocomposite-based strain sensor.

## The analytical model and calculating formulas of the M-NW nanocomposite based interdigital strain sensor.

As presented by the cross-sectional schematic diagram of the interdigital strain sensor and the layout of the interdigitated electrode. The gaps width between neighboring electrode fingers is g, and the width and length of electrode fingers are w and L, respectively. The metallization ratio of the interdigitated electrode is demonstrated by the equation:

$$\eta = \frac{w}{w+g} \qquad (4)$$

According to the electric potential boundary planes distribution for cross-sectional periodic electrode fingers. The total capacitance of the interdigital strain sensor is calculated by the follow equation:

$$C = (N-3)\frac{C_I}{2} + \frac{2C_E C_I}{C_E + C_I}, \text{ for } N > 3 \qquad (5)$$

where $N$ is the number of electrode fingers; $C_I$ is half the capacitance of one interior electrode relative to the ground potential, and $C_E$ is the capacitance of one outer electrode relative to the ground plane. $C_I$ and $C_E$ are computed with the modulus as follows:

$$C_I = \varepsilon_0 \varepsilon_r L \frac{K(k)}{K(k')} \qquad (6)$$

$$C_E = \varepsilon_0 \varepsilon_r L \frac{K(k_E)}{K(k'_E)} \qquad (7)$$

$K(k)$ is the elliptic integral of the first type with the modulus $k$, and the complementary modulus $k'$. $\varepsilon_r$ and $\varepsilon_0$ represent the relative dielectric permittivity and free space. The modulus $k$ is

$$k = \sin\left(\frac{\pi}{2}\eta\right) \qquad (8)$$

$$k_E = \frac{2\sqrt{\eta}}{1+\eta} \qquad (9)$$

and the complementary modulus $k'$ is

$$k' = \sqrt{1-k^2} \qquad (10)$$

$$k'_E = \sqrt{1-k_E^2} \qquad (11)$$

According to the above equations, the total capacitance of the interdigital capacitive strain sensor is given by

$$C = \varepsilon_0 \varepsilon_r L \left( \frac{(N-3)}{2} \frac{K(k)}{K(k')} + \frac{2K(k_E)K(k)}{K(k_E)K(k') + K(k'_E)K(k)} \right) \quad (12)$$

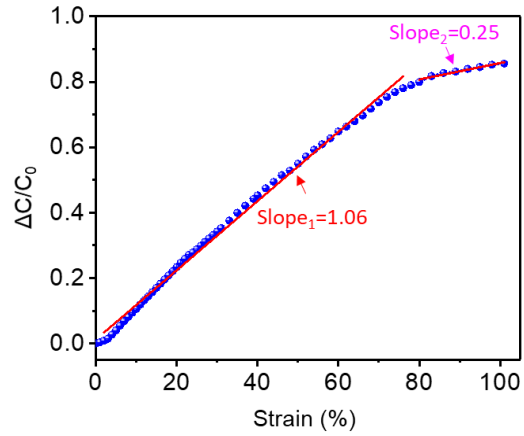

**Figure S11.** Sensitivity curve of capacitance change of M-NW nanocomposite-based interdigital strain sensor under longitudinal strain.

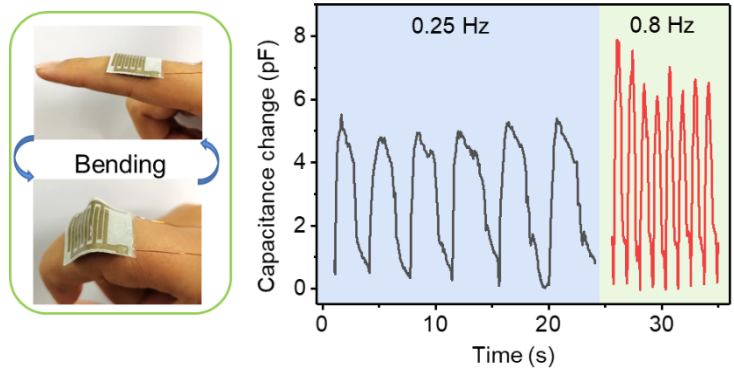

**Figure S12.** The relative change in capacitance response of the interdigital strain sensor under bend–release finger motions at different frequency.

**Table S1.** The performance comparisons among elastic dielectric nanocomposites.

| Materials | Inclusion | Strategies | Fillers content | Dielectric permittivity | Ref. |
|---|---|---|---|---|---|
| PDMS | BaTiO$_3$ nanoparticles | Composite | 26 vt% | 9 @100 Hz | 1 |
| PDMS | TiO$_2$ nanotubes | Composite | 5 wt% | 4.4 @1 kHz | 2 |
| PDMS | Polyphenylmethylsiloxane | Cross-linking synthesis | ⸺ | 3.9 @1 MHz | 3 |
| SBAS | Graphene oxide | Composite | 2 wt% | 58 @1 kHz | 4 |
| Cellulose nanofibers-epoxy | AgNFs | Composite | 1.5 wt% | 9.2 @120 kHz | 5 |
| Acrylate copolymer | $Al_2O_3$ | Composite | 4 vt% | 8.5 @100 Hz | 6 |
| Acrylate | Glycidyl methacrylate | Synthesis | —— | 5.67 @1 kHz | 7 |
| PU | $BaTiO_3$ nanoparticles | Composite | 50 phr | 13.6 @1 kHz | 8 |
| TPU | Silver nanowires (AgNW) | Composite | 1 vt% | 13 @1 kHz | 9 |
| TPU | Carbon nanospheres | Composite | 3 wt% | 11 @1 kHz | 10 |
| TPU | $BaTiO_3$ nanoparticles | Composite | 30 vol% | 31 @1 kHz | 11 |
| TPU | $BaTiO_3$ nanoparticles AgNW | Nanonetwork & multi-layers | 11 vol% BTO, 0.24 vol% AgNW | 113.36 @ 1 kHz | **This work** |

**Table S2.** The performance comparisons among elastic dielectric nanocomposites.

| Samples | Conductivity | Stretchability | Young's modulus (MPa) | $\varepsilon_r$ (1kHz) | $\tan\theta$ (1 kHz) | Electromechanical sensitivity (Mpa$^{-1}$) $\beta=\varepsilon_r/Y$ | Comprehensive Electromechanical Performance (Mpa$^{-1}$) FOM =$\varepsilon_r$/Y $\tan\theta$ |
|---|---|---|---|---|---|---|---|
| TPU | 2.69×10$^{-10}$ | 514% | 4.0 | 7.34 | 0.031 | 1.84 | 59.19 |
| P-0-3 | 2.20×10$^{-10}$ | 291% | 6.7 | 16.77 | 0.028 | 2.50 | 89.39 |
| S-NW | 4.42×10$^{-10}$ | 260% | 4.8 | 29.98 | 0.034 | 6.25 | 183.70 |
| M-NW | 1.94×10$^{-9}$ | 360% | 7.2 | 113.36 | 0.029 | 15.74 | 542.91 |

**Table S3.** The performance comparisons among reported different capacitive strain sensors.

| Dielectric materials | Electrodes | Type of sensors | Gauge factor | linearity range | Ref. |
|---|---|---|---|---|---|
| PDMS | SWCNT | Parallel-plate capacitive | 0.4 | 50% | 1 |
| Silicone elastomer | Aluminum & silver | Parallel-plate capacitive | 0.9 | 85% | 2 |
| Acrylic elastomer | AgNW-PU composite | Parallel-plate capacitive | 0.5 | 60% | 3 |
| Ecoflex | AgNW-PDMS | Parallel-plate capacitive | 0.7 | 50% | 4 |
| Silicone | SWCNT | Parallel-plate capacitive | 0.99 | 100% | 5 |
| Ecoflex | Carbon black-elastomer composite | Parallel-plate capacitive | 0.83-0.98 | 500% | 6 |
| Multidimensional carbon/polyurethane | Gold | Parallel-plate capacitive | 0.45 | 100% | 7 |
| Ecoflex | Conductive knit fabric | Parallel-plate capacitive | 1.35 | 100% | 8 |
| PDMS-MPU$_{0.4}$-IU$_{0.6}$ | CNT/elastic polymer | Parallel-plate capacitive | 0.24 | 50% | 9 |
| VHB | MXene/PVA | Parallel-plate capacitive | 0.4 | 200% | 10 |
| PDMS | Vertical graphene | Parallel-plate capacitive | 0.97 | 80% | 11 |
| Rubber | Gold | Interdigital capacitive | -0.98 | 14% | 12 |
| PDMS | AgNW | Interdigital capacitive | -2.0 | 30% | 13 |
| BTO-Ecoflex | Carbon black-Ecoflex | Interdigital capacitive | -0.45 | 100% | 14 |
| Nanonetwork & multi-layer nanocomposite | TPU-AgNW | Interdigital capacitive | 1.06 | 100% | **This work** |